\documentclass[a4paper]{jpconf}
\newcommand{\bea}{\begin{eqnarray}}
\newcommand{\beq}{\begin{equation}}
\newcommand{\eea}{\end{eqnarray}}
\newcommand{\eeq}{\end{equation}}
\usepackage{graphicx}
\begin{document}
\title{A geometrical crossover in excited states of
two-electron quantum dots in a magnetic field}

\author{R.G.  Nazmitdinov $^{1,2}$,  N. S. Simonovi\'c $^{3}$, 
A. R. Plastino $^{4,5}$, and A. V. Chizhov $^{2}$}

\address{$^{1}$ Departament de F{\'\i}sica, Universitat de les Illes Balears, 
07122 Palma de Mallorca, Spain}

\address{$^{2}$ BLTP, Joint Institute for Nuclear Research, 141980 Dubna, Russia }

\address{$^{3}$ Institute of Physics, University of Belgrade, 11001 Belgrade, Serbia }

\address{$^{4}$ Instituto Carlos I de Fisica Teorica y Computacional, Universidad de Granada, 18071 Granada, Spain }

\address{$^{5}$ National University La Plata, UNLP-CREG-CONICET, C.C. 727, La Plata 1900, Argentina }

\ead{rashid@theor.jinr.ru}

\begin{abstract}
We use the entanglement measure to study  the evolution of quantum correlations in
two-electron axially-symmetric parabolic quantum dots under a
perpendicular magnetic field. 
We found that the entanglement indicates on the shape transition
in the density distribution of two electrons in 
the lowest state with zero angular momentum projection 
at the specific value of the applied magnetic field.
\end{abstract}

\section{Introduction}

Nowadays there is a growing interest in using quantum entanglement
measures for study of quantum correlations in topologically
ordered systems \cite{2}. The analysis of these systems is a
highly non-trivial task due to the absence of the order parameter.
Particular examples are the integer and fractional quantum Hall
liquids  which cannot be understood in terms of the traditional
description of phases based on symmetry breaking and
local order parameters. 
The main stream of such an analysis is focused on many-particle
one-dimensional systems (cf \cite{kit}). Recently, topological
phases were studied in finite systems such as two-dimensional
electrons at very high magnetic field (Laughlin states) \cite{zoz}
and weakly interacting two-dimensional rotating Bose-Einstein
condensate \cite{ved}. 
There are attempts to find 
a relation between  topological orders 
of different topological phases and quantum phase transitions,
driven by quantum fluctuations at zero temperature in many-body systems 
(see a textbook \cite{Sad}). 
 
It is believed that various quantum
phases could exist in quantum dots (QDs) at different strengths
of the applied perpendicular magnetic field \cite{RM}. 
At small magnetic field strengths one observes 
the orbital momentum and spin
oscillations of the ground state of a QD by increasing the field
strength.
At certain field range the oscillations disappear and  electrons
form a fully polarized state called the maximum density droplet.
It is expected that a further increase of the magnetic field  
should lead to the formation of the Wigner molecule, a finite-size analogue of
the Wigner crystallization of the homogeneous electron gas. 
A natural question arises: if QDs can be considered as a finite-size
analogy of conventional condensed matter systems what are  
signatures of quantum phase transitions in QDs ? 

To shed light on this question we will employ the 
entanglement and compare its evolution with the evolution of  quantum 
spectra of QDs as a function of the magnetic field.
Evidently, finite systems can only show precursors of the  QPT
behaviour. However, they are also important for the development of the concept.
Two-electron QDs being realictic tractable nontrivial systems
are, in particular, attractive because their eigenstates can be obtained 
very accurately, or in some cases, exactly (cf \cite{taut,r1}). 
Moreover, it was found that at certain values of the magnetic field 
quantum spectra of two-electron QDs become degenerate due
to onset of the spherical symmetry \cite{hs,hs1}.
The goal objective of the present
paper is to demonstrate that a quantum entanglement can be used to
indicate this transition in a three-dimensional (3D) quantum dot under a magnetic
field.

\section{Basics}

Our analysis is carried out by means of the numerical diagonalization
of the Hamiltonian \beq \label{ham} H = \sum_{j=1}^2 \bigg[
\frac{1}{2m^*\!}\, \Big({\bf p}_j - \frac{e}{c} {\mathbf A}_j
\Big)^{\! 2} + U({\mathbf r}_j) \bigg] + \frac{k}{\vert{\mathbf
r}_1 \!- {\mathbf r}_2\vert}+ H_{\it spin}. \eeq Here $k =
e^2/4\pi\varepsilon_0\varepsilon_r$ and $H_{\it
spin}=g^*\mu_B(\mathbf{s}_1+\mathbf{s}_2)\!\cdot\!\mathbf{B}$
describes the Zeeman term, where $\mu_B=e\hbar/2m_ec$ is the
Bohr magneton. As an example, we will use 
the effective mass $m^*=0.067m_e$, the relative
dielectric constant $\varepsilon_r=12$ and the effective Land\'e
factor $g^*=-0.44$ (bulk GaAs values). 
For the perpendicular magnetic field we choose the vector
potential with gauge ${\mathbf A} = \frac{1}{2} \mathbf{B} \times
\mathbf{r} = \frac{1}{2}B(-y, x,0)$. The confining potential is
approximated by a 3D axially-symmetric harmonic oscillator
$U(\mathbf{r}) = m^* [\omega_0^2\,(x^2 \!+ y^2) +
\omega_z^2z^2]/2$, where $\hbar\omega_z$ and $\hbar\omega_0$ are
the energy scales of confinement in the $z$-direction and in the
$xy$-plane, respectively.

By introducing the center of mass (CM) and relative coordinates:
$\mathbf{R} = \frac{1}{2}(\mathbf{r}_1 + \mathbf{r}_2)$ and
$\mathbf{r}_{12} = \mathbf{r}_1 - \mathbf{r}_2$, --  the
Hamiltonian~(\ref{ham}), in agreement with the Kohn theorem \cite{kohn},
separates into the CM and relative-motion terms
$H = H_{\rm CM} + H_{\rm rel}$ (see details in \cite{r1}).
The CM term is described by
 the oscillator Hamiltonian with the mass ${\cal M} = 2m^*$ and
frequencies  of the one-particle confining potential $U$.
The Hamiltonian for relative motion
in cylindrical coordinates takes the form
\begin{equation}
H_{\rm rel} = \frac{1}{2\mu}\Big(p_{\rho_{12}}^2 \!+
\frac{l_z^2}{\rho_{12}^2} + p_{z_{12}}^2\!\Big) +
\frac{\mu}{2}(\Omega^2 \rho_{12}^2 + \omega_z^2 z_{12}^2) +
\frac{k}{r_{12}} - \omega_L l_z, \label{relham}
\end{equation}
where $\mu = m^*/2$ is the reduced mass, $l_z$ ($\to
-i\hbar\partial/\partial\varphi_{12}$) is the projection of
angular momentum for relative motion and $\rho_{12} = (x_{12}^2 +
y_{12}^2)^{1/2}$, $\varphi_{12} = \arctan(y_{12}/x_{12})$, $r_{12}
= (\rho_{12}^2 + z_{12}^2)^{1/2}$. Here, $\omega_L = eB/2m^*\!c$
is the Larmor frequency, and the effective lateral confinement
frequency $\Omega = (\omega_{L}^{2} +
\omega_{0}^{2})^{1/2}$ depends through $\omega_{L}$ on the magnetic
field.

The total two-electron wave function
$\Psi(\mathbf{r}_1,\mathbf{r}_2) =
\psi(\mathbf{r}_1,\mathbf{r}_2)\chi(\sigma_1,\sigma_2)$ is a
product of the orbital $\psi(\mathbf{r}_1,\mathbf{r}_2)$ and spin
$\chi(\sigma_1,\sigma_2)$ wave functions.
Due to the Kohn theorem, the orbital wave function is factorized
as a product of the CM and the relative motion wave functions
\begin{equation}
\psi(\mathbf{r}_1,\mathbf{r}_2) =
\psi_\mathrm{CM}(\mathbf{R})\,\psi_\mathrm{rel}(\mathbf{r}_{12}).
\label{psiorb}
\end{equation}

According to the Pauli principle, the orbital wave function must
be symmetric (or, equivalently, $\psi_\mathrm{rel}
(\mathbf{r}_{12})$ must be even) for the antisymmetric (singlet:
$S = M_S = 0$) spin state, and it must be antisymmetric
($\psi_\mathrm{rel}(\mathbf{r}_{12})$ must be odd) for the
symmetric (triplet: $S = 1$, $M_S = 0,\pm1$) spin states. Thus,
for the relative motion the parity of $\psi_\mathrm{rel}
(\mathbf{r}_{12})$ is a good quantum number as well as the
magnetic quantum number $m$, since $l_z$ is the integral of
motion.

The CM eigenfunction is a product of the Fock-Darwin state (the
eigenstate of electron in an isotropic 2D harmonic oscillator potential
under a perpendicular magnetic field) \cite{Fock} in the
$(X,Y)$-plane and the oscillator function in the $Z$-direction
(both sets for a particle of mass ${\cal M}$). In this paper we
consider the lowest CM eigenstate which has the form
$\psi_\mathrm{CM}(\mathbf{R}) =\psi^{(xy)}_\mathrm{CM}(X,Y)\,
\psi^{(z)}_\mathrm{CM}(Z)$, where $\psi^{(xy)}_\mathrm{CM}(X,Y) =
\sqrt{2\bar{\Omega}/\pi}\, e^{-\bar{\Omega}(X^2 + Y^2)}$ and
$\psi^{(z)}_\mathrm{CM}(Z) = (2\bar{\omega}_z/\pi)^{1/4} e^{-
\bar{\omega}_z Z^2}$ (i.e. zero principle quantum numbers), with
$\bar{\Omega} = m^* \Omega/\hbar$ and $\bar{\omega}_z = m^*
\omega_z/\hbar$.

Since the Coulomb interaction mixes the eigenstates of
non-interacting electrons, the eigenfunctions of the Hamiltonian
for relative motion (\ref{relham}) are expanded in the basis of
the Fock-Darwin states $\Phi_{n,m}(\rho_{12}, \varphi_{12})$ and
oscillator functions in the $z_{12}$-direction
$\phi_{n_z}(z_{12})$ (for a particle of mass $\mu$), i.e.
\begin{equation}
\psi_\mathrm{rel}(\mathbf{r}_{12}) = \sum_{n,n_z} c_{n,n_z}^{(m)}
\Phi_{n,m}(\rho_{12},\varphi_{12})\,\phi_{n_z}(z_{12}).
\label{expansion_3d}
\end{equation}
The coefficients $c_{n,n_z}^{(m)}$ can be determined by
diagonalizing the Hamiltonian (\ref{relham}) in the same basis.
Evidently, in numerical analysis the basis is restricted to a
finite set $\{\Phi_{n,m}\,\phi_{n_z} |\, n = 0,\ldots,n_{\max};\,
n_z = 0,\ldots,n_z^{\max}\}$. It  must be, however, large enough
to provide a good convergence for the numerical results. Since the
function $\psi_\mathrm{rel} (\mathbf{r}_{12})$ has a definite
parity and the parity of the functions $\Phi_{n,m}\, \phi_{n_z}$
is $(-1)^{m + n_z}$, the index $n_z$ in the expansion
(\ref{expansion_3d}) takes either even or odd values.

For non-interacting electrons ($k = 0$) the eigenfunctions
$\psi_\mathrm{rel}$ are simply the basis functions
$\Phi_{n,m}\,\phi_{n_z}$, and, therefore, the ground state is
described by the wave function $\psi_\mathrm{rel} =
\Phi_{0,0}\,\phi_0$.
When two interacting electron move in the external field created
by the confining potential and the applied, varying steadily,
magnetic field, the quantum number $m$ of the ground state 
(in the form (\ref{expansion_3d})) evolves 
from zero to higher values as the  magnetic field strength
increases. It results in the well known singlet-triplet
transitions \cite{wag}.
Namely, for a given $m$ the dominant term in the expansion
(\ref{expansion_3d}) will be $\Phi_{0,m}\,\phi_0$ ($\Rightarrow$
all $n_z$ are even) and the parity of the ground state is
$(-1)^m$, which determines the total spin to be $S = \frac{1}{2}[1
- (-1)^m]$. Note that the quantum number $M_S$ associated with the
spin wave function evolves as following: for even $m$ the total
spin $S = 0$ and, thus, $M_S = 0$; for odd $m$  the total spin $S
= 1$ and $M_S$ can be $-1$, $0$ or $1$. The Zeeman splitting (with
$g^* < 0$) will lower the energy of the $M_S =1$ component of the
triplet states, while leaving the singlet states unchanged. As a
consequence, the ground state will be characterized by $M_S = S$.
With the  increasing magnetic field the intervals of the triplet
states will increase at the cost of the singlet ones, and
eventually, the singlet ground states will be totally suppressed.
The increase of the magnetic field leads to the formation of a
ring and a torus of  maximal density in 2D- and 3D-densities,
respectively (see Fig.~4 in Ref.\cite{nen3}).

At the value $\omega_L^\mathrm{sph} = (\omega_z^2 -
\omega_0^2)^{1/2}$ the magnetic field gives rise to the {\it
spherical symmetry} $(\omega_z/\Omega=1)$ (with $\omega_z >
\omega_0$)  in the {\it axially-symmetric}  two-electron QD
\cite{hs,hs1}. This phenomenon was also recognized in the results
for many interacting electrons in self-assembled QDs \cite{W}. 
In the later case it was interpreted as an approximate symmetry that had
survived from the non-interacting case due to the dominance of the
confinement energy over a relatively small Coulomb interaction
energy. However, the symmetry is not approximate but {\it exact}
even for strongly interacting electrons, because the radial
electron-electron repulsion does not break the rotational
symmetry. A natural question arises how to detect such a
transition looking on the density distribution only. The related
question is, if such a transition occurs, what are 
the concomitant structural  changes?

To this end we employ  the entanglement measure based on the
linear entropy of reduced density matrices (cf \cite{yuk})
\begin{equation}
{\cal E} = 1 - 2\,\mathrm{Tr}[{\rho_r^{(orb)}}^2]\,
\mathrm{Tr}[{\rho_r^{(spin)}}^2], \label{ent_measure}
\end{equation}
where $\rho_r^{(orb)}$ and $\rho_r^{(spin)}$ are the 
single-particle reduced density matrices in the orbital and spin spaces,
respectively. This measure is quite popular for the analysis of the entanglement
of two-fermion systems, in particular, two electrons confined in the parabolic 
potential in the absence of the magnetic field \cite{YPD}.

The trace $\mathrm{Tr}[{\rho_r^{(spin)}}^2]$ of the two-electron
spin states with a definite symmetry $\chi_{S,M_S}$ has two
values: (i) $1/2$ if $M_S = 0$ (anti-parallel spins of two
electrons); (ii) $1$ if $M_S = \pm1$ (parallel spins). The
condition $M_S = S= \frac{1}{2}[1- (-1)^m]$ yields
\begin{equation}
\mathrm{Tr}[{\rho_r^{(spin)}}^2] = \hbox{$\frac{1}{2}$}(1 +
|M_S|)=\frac{3-(-1)^m}{4}.
\label{trace_spin}
\end{equation}

The trace of the orbital part $\mathrm{Tr}[{\rho_r^{(orb)}}^2]$
\begin{eqnarray}
\mathrm{Tr}[{\rho_r^{(orb)}}^2] \!\!&=&\!\! \int d\mathbf{r}_1\,
d\mathbf{r}_1^{\,\,\prime}\, d\mathbf{r}_2\,
d\mathbf{r}_2^{\,\,\prime}\, \psi(\mathbf{r}_1,\mathbf{r}_2)\,
\psi^*(\mathbf{r}_1^{\,\,\prime},\mathbf{r}_2) \nonumber
\\[-.5ex]
&&\qquad\quad \psi^*(\mathbf{r}_1,\mathbf{r}_2^{\,\,\prime})\,
\psi(\mathbf{r}_1^{\,\,\prime},\mathbf{r}_2^{\,\,\prime}).
\end{eqnarray}
is more involved. Indeed, in virtue of Eqs.(\ref{psiorb}), (\ref{expansion_3d}),
one obtains
\begin{eqnarray}
\label{gen}
\mathrm{Tr}[{\rho_r^{(orb)}}^2] \!\!\!&=&\!\!\!
\sum_{n_1=0}^{n_{\max}} \sum_{n_2=0}^{n_{\max}}
\sum_{n_3=0}^{n_{\max}} \sum_{n_4=0}^{n_{\max}}
\sum_{n_{z_1\!}=0}^{n_z^{\max}} \sum_{n_{z_2}\!=0}^{n_z^{\max}}
\sum_{n_{z_3}\!=0}^{n_z^{\max}} \sum_{n_{z_4}\!=0}^{n_z^{\max}}
\nonumber
\\
&&\qquad c_{n_1,n_{z_1}}^{(m)} c_{n_2,n_{z_2}}^{(m)}
c_{n_3,n_{z_3}}^{(m)} c_{n_4,n_{z_4}}^{(m)} \label{trace3d}
\\
&& I(n_1,n_2,n_3,n_4;m)\, J(n_{z_1},n_{z_2},n_{z_3},n_{z_4}),
\nonumber
\end{eqnarray}
where
\begin{eqnarray}
&&I(n_1,n_2,n_3,n_4;m) = \int d\mathbf{r}_1\,
d\mathbf{r}_1^{\,\prime}\, d\mathbf{r}_2\,
d\mathbf{r}_2^{\,\prime}
\\
&&\psi^{(xy)}_\mathrm{CM}(\hbox{$\frac{\mathbf{r}_1 +
\mathbf{r}_2}{2}$})\,
{\psi_\mathrm{CM}^{(xy)}}^{\!*}(\hbox{$\frac{\mathbf{r}_1^{\,\prime}
+ \mathbf{r}_2}{2}$})\,
{\psi_\mathrm{CM}^{(xy)}}^{\!*}(\hbox{$\frac{\mathbf{r}_1 +
\mathbf{r}_2^{\,\prime}}{2}$})\,
\psi_\mathrm{CM}(\hbox{$\frac{\mathbf{r}_1^{\,\prime} +
\mathbf{r}_2^{\,\prime}}{2}$})\nonumber
\\
&&\Phi_{n_1,m}(\mathbf{r}_1 \!-\! \mathbf{r}_2)\,
\Phi_{n_2,m}^*(\mathbf{r}_1^{\,\prime} \!-\! \mathbf{r}_2)\,
\Phi_{n_3,m}^*(\mathbf{r}_1 \!-\!\mathbf{r}_2^{\,\prime})\,
\Phi_{n_4,m}(\mathbf{r}_1^{\,\prime} \!-\!
\mathbf{r}_2^{\,\prime}) \nonumber
\end{eqnarray}
(here $\mathbf{r}_i$ are vectors in the $xy$-plane) and
\begin{eqnarray}
&&J(n_{z_1},n_{z_2},n_{z_3},n_{z_4}) =\int dz_1\,
dz_1^{\,\prime}\, dz_2\, dz_2^{\,\prime}
\\
&&\psi_\mathrm{CM}^{(z)}(\hbox{$\frac{z_1 + z_2}{2}$})\,
{\psi_\mathrm{CM}^{(z)}}^{\!\!*}(\hbox{$\frac{z_1^{\,\prime} +
z_2}{2}$})\, {\psi_\mathrm{CM}^{(z)}}^{\!\!*}(\hbox{$\frac{z_1 +
z_2^{\,\prime}}{2}$})\,
\psi_\mathrm{CM}^{(z)}(\hbox{$\frac{z_1^{\,\prime} +
z_2^{\,\prime}}{2}$})\nonumber
\\
&&\phi_{n_{z_1}}(z_1 \!-\! z_2)\, \phi_{n_{z_2}}^*(z_1^{\,\prime}
\!-\! z_2)\, \phi_{n_{z_3}}^*(z_1 \!-\! z_2^{\,\prime})\,
\phi_{n_{z_4}}(z_1^{\,\prime} \!-\! z_2^{\,\prime}). \nonumber
\end{eqnarray}

The magnetic field dependence of the entanglement ${\cal E}$
naturally occurs via inherent variability of the expansion
coefficients. The values of the $I$ and $J$ integrals for any
choice of indices can be determined analytically,
which simplifies the numerical calculations.

\begin{figure}[h]
\begin{center}
\includegraphics[width=14pc]{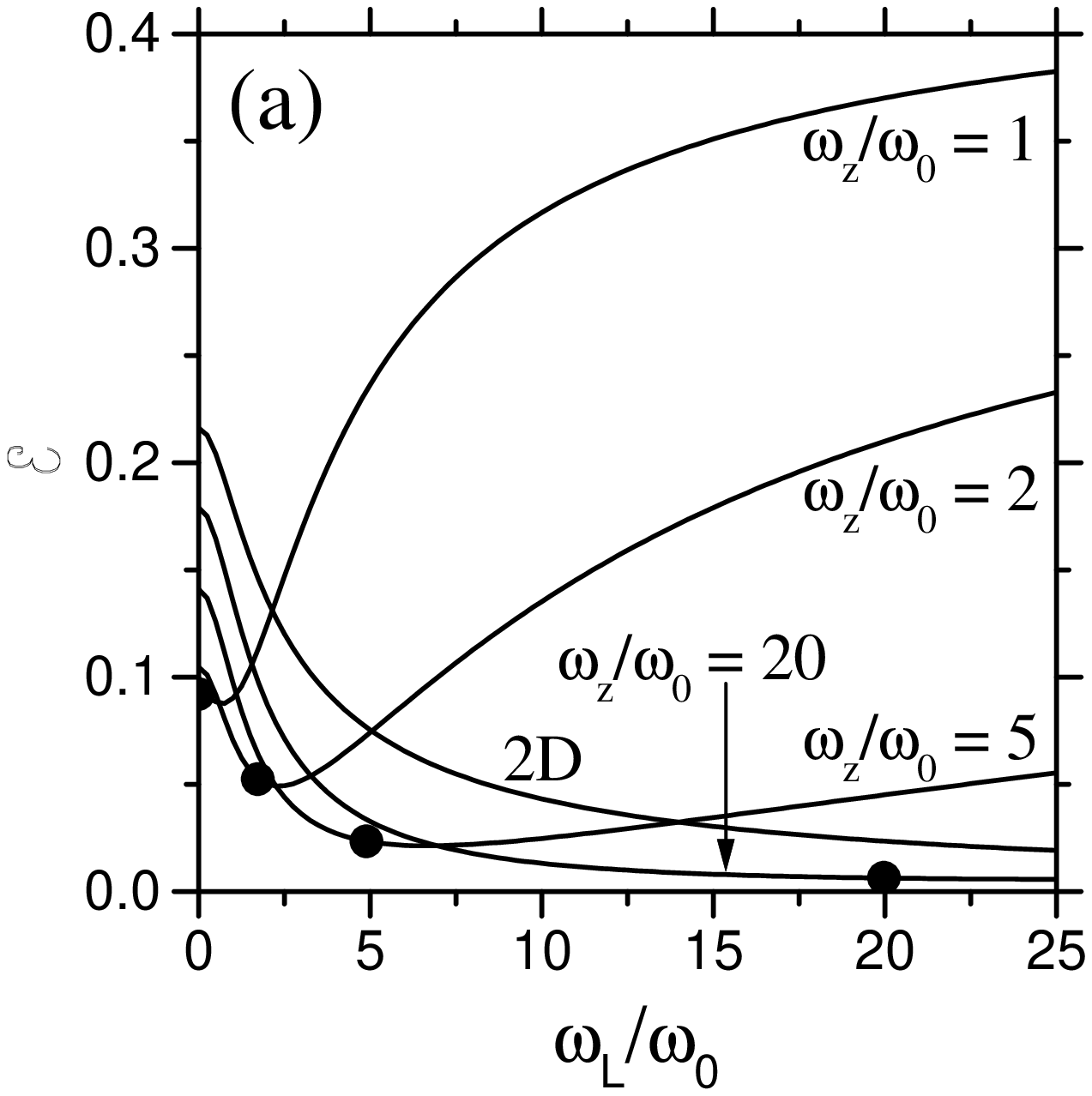}\hspace{2pc}%
\includegraphics[width=14pc]{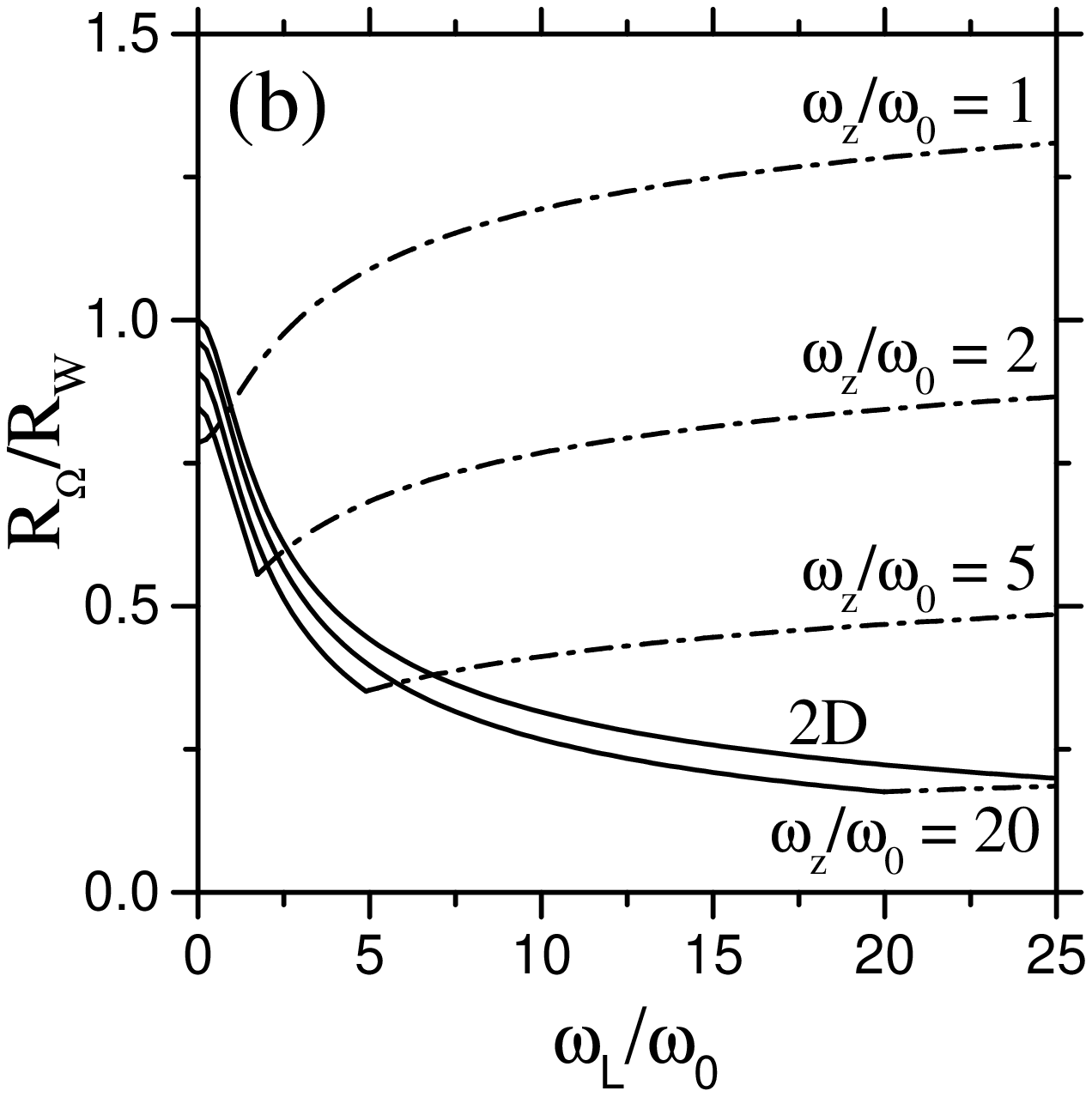}
\end{center}
\caption{\label{ent-oml}
(a) Entanglement of the lowest state with $m = 0$ at $R_W
= 2$ and various ratios $\omega_z/\omega_0$ as functions of the
parameter $\omega_L/\omega_0$. The circles denote the values
of $\omega_L/\omega_0$ when QDs with given ratios
$\omega_z/\omega_0$ become spherically symmetric. (b) The relative
strength of the Coulomb interaction $R_\Omega^\mathrm{(2D)}/R_W$
(solid line) and $R_\Omega^\mathrm{(1D)}/R_W$ (dash-dotted line)
for the lowest state with $m = 0$ at various ratios
$\omega_z/\omega_0$ as functions of the parameter
$\omega_L/\omega_0$.} 
\end{figure}

For our analysis it is convenient to use the so-called Wigner
parameter $R_W = (k/l_0)/\hbar\omega_0 = l_0/a^*$, 
a measure of the  Coulomb interaction strength relative to the  confinement
strength (cf \cite{r1}). Here, $l_0 = \sqrt{\hbar/m^*\omega_0}$ is the oscillator
length and $a^* = \hbar^2/km^*$ is the effective Bohr radius. 
For our choice of the parameters (GaAs) and for the confinement
frequency $\hbar\omega_0\approx 2.8$ meV we have $R_W\approx 2$.
The numerical analysis demonstrates a good convergency for the
basis with $n_{\it max}=n_z^{\it max}=4$.

\section{Analysis of results}

The strongest correlations are expected for the states with $m=0$, since they are
not affected by the orbital electron motion which weakens the Coulomb interaction.
In the absence of the magnetic field ($B = 0$) the entanglement
decreases if the ratio $\omega_z/\omega_0$ decreases from $\infty$
(2D model) to $1$ (spherically symmetric 3D model); see
Fig.~\ref{ent-oml}(a) at $\omega_L/\omega_0 = 0$. This effect
could be explained by introducing the effective charge
$k_\mathrm{eff}$ \cite{hs,ec} which determines the effective
electron-electron interaction $V_C=k_\mathrm{eff}/\rho_{12}$ in
the QD. In the 3D dot the electrons can avoid each other more
effectively than in the 2D one. Therefore, the
Coulomb interaction has a smaller effect on the 3D spectrum (the
ratio $k_\mathrm{eff}/k \sim 0.5$) in contrast to the 2D case when
$k_\mathrm{eff}/k = 1$. Thus, a decreasing of the ratio
$\omega_z/\omega_0$ yields an analogous effect as the reduction of
the electron-electron interaction.

Fig.~\ref{ent-oml}(a) shows the entanglement measure $\cal E$ of
the lowest angular momentum state $m = 0$ as a function of the
magnetic field (the parameter $\omega_L/\omega_0$) at a fixed
value of $R_W$ and for different ratios $\omega_z/\omega_0$. In
the 2D case the entanglement decreases monotonically with the
increase of the magnetic field. The constant electron-electron
interaction becomes relatively weaker, since the effective lateral
confinement ($\hbar\Omega$) increases with the magnetic field. If
we introduce the characteristic length of the effective
confinement $l_\Omega = \sqrt{\hbar/m^*\Omega}$, the parameter
$R_\Omega = l_\Omega/a^*$ (which is equal $R_W$ at $B = 0$)
determines the relative strength of Coulomb interaction at a given
effective confinement. Evidently, $R_\Omega$ decreases with the
increase of the magnetic field $B$ (see Fig.~\ref{ent-oml}(b), the
line labelled by '2D'). In the 3D case, however, the entanglement
decreases until $\omega_L = \omega_L^\mathrm{sph}$, when the
spherical symmetry occurs. After this point the entanglement
starts to increase (see Fig.~\ref{ent-oml}(a)).

This behaviour can be explained by the influence of magnetic field
on the effective strength $R_\Omega$, which is  twofold here.
Indeed, in the 3D case the magnetic field affects the effective
charge as well as the effective confinement. For the quasi-2D
system of electrons ($\Omega \ll \omega_z$)  the effective charge
is $k_\mathrm{eff}^\mathrm{(2D)} = \langle\rho_{12}V_C\rangle$
(see Eq.~(18) in Ref. \cite{ec}), where $V_C = k/\sqrt{\rho_{12}^2
+ z_{12}^2}$ is the full 3D Coulomb interaction. The mean value
$\langle\rho_{12}V_C\rangle$ is calculated by means of the
eigenstates of $H_\mathrm{rel}$ in the approximation of
non-interacting electrons. Here, the eigenstate is $\Phi_{0,m}\,
\phi_0$ (for explicit expressions see Eqs.~(19),(20) in Ref.
\cite{ec}). Thus, for the quasi-2D case the parameter
$R_\Omega^\mathrm{(2D)} = (m^*/\hbar^3\Omega)^{1/2}\,
k_\mathrm{eff}^\mathrm{(2D)}$ can be used as a measure for the
relative strength of the Coulomb interaction.

For $\Omega \gg \omega_z$ (very strong magnetic field) the
electrons are pushed laterally towards the dot's center. The
magnetic field, however, does not affect the vertical confinement.
As a consequence the electrons practically can move only in the
z-direction and the QD becomes a quasi-1D system. In this case a
measure for the relative strength of Coulomb interaction can be
defined as $R_\Omega^\mathrm{(1D)} = (m^*/\hbar^3\omega_z)^{1/2}\,
k_\mathrm{eff}^\mathrm{(1D)}$, where the effective charge for a
quasi-1D system is $k_\mathrm{eff}^\mathrm{(1D)} = \langle
|z_{12}|V_C\rangle$. It can be shown  that for the lowest state
with $m = 0$ one obtains $k_\mathrm{eff}^\mathrm{(1D)}/k = (1
+\sqrt{\omega_z/\Omega})^{-1}$.

The quantities $R_\Omega^\mathrm{(2D)}$ and
$R_\Omega^\mathrm{(1D)}$ for the lowest state with $m = 0$, as
functions of the parameter $\omega_L/\omega_0$ (in the domains $0
< \omega_L < \omega_L^\mathrm{sph}$ and $\omega_L >
\omega_L^\mathrm{sph}$, respectively), are shown in
Fig.~\ref{ent-oml}(b) for different ratios $\omega_z/\omega_0$.
One observes that the effective strength $R_\Omega^\mathrm{(2D)}$
decreases with the increase of the magnetic field for different
ratios $\omega_z/\omega_0$, similar to the 2D case. The oppositely
ordered confinement $\Omega^\mathrm{(1D)}$ (which is not defined
for the 2D case) increases with $\omega_L$ and, therefore, the
effective strength $R_\Omega^\mathrm{(1D)}$ increases as well. In
order to match $R_\Omega^\mathrm{(1D)} = R_\Omega^\mathrm{(2D)}$
at $\omega_L = \omega_L^\mathrm{sph}$ (i.e. when $\Omega =
\omega_z$) the strength $R_\Omega^\mathrm{(1D)}$ is scaled by the
factor $\pi/2$. Although at this point the 3D system is far  from
the 2D model and from the 1D model and, as a consequence,
$R_\Omega^\mathrm{(2D)}$ and $R_\Omega^\mathrm{(1D)}$ do not match
smoothly, these two functions taken together give a qualitative
picture how the effective electron-electron interaction in a 3D QD
changes with the magnetic field.

The minimum of entanglement for the lowest state with $m=0$ 
at $\Omega = \omega_z$ can be associated
with the condition of equivalence of the oscillator scales in
the lateral and vertical confinements: $l_\Omega \equiv l_z =
\sqrt{\hbar/m^*\omega_z}$. These quantities can be understood as
the amplitudes of electron oscillations in the $\rho$ and
$z$-directions, respectively, for single-particle states with
$n_\rho = 0$ and $n_z = 0$ \cite{ec}. At this
point the effective Coulomb interaction becomes
isotropic, which results in the small mixing of single-particle
states, in contrast to the 2D and 1D cases. This result can be
readily extended for the infinite square well potential in the
z-direction. In this case $l_z\approx d/2$; $d$ is the thickness
of semiconductor layer where the dot is created (see Fig.1 in
Ref.~\cite{ec})). Therefore, the minimum of entanglement is
expected at $l_\Omega = d/2$ which defines the value of the
magnetic field by means of the Larmor frequency $\omega_L =
\sqrt{(4\hbar/m^*d^2)^2-\omega_0^2}$.

\begin{figure}
\begin{center}
\includegraphics[width=21pc]{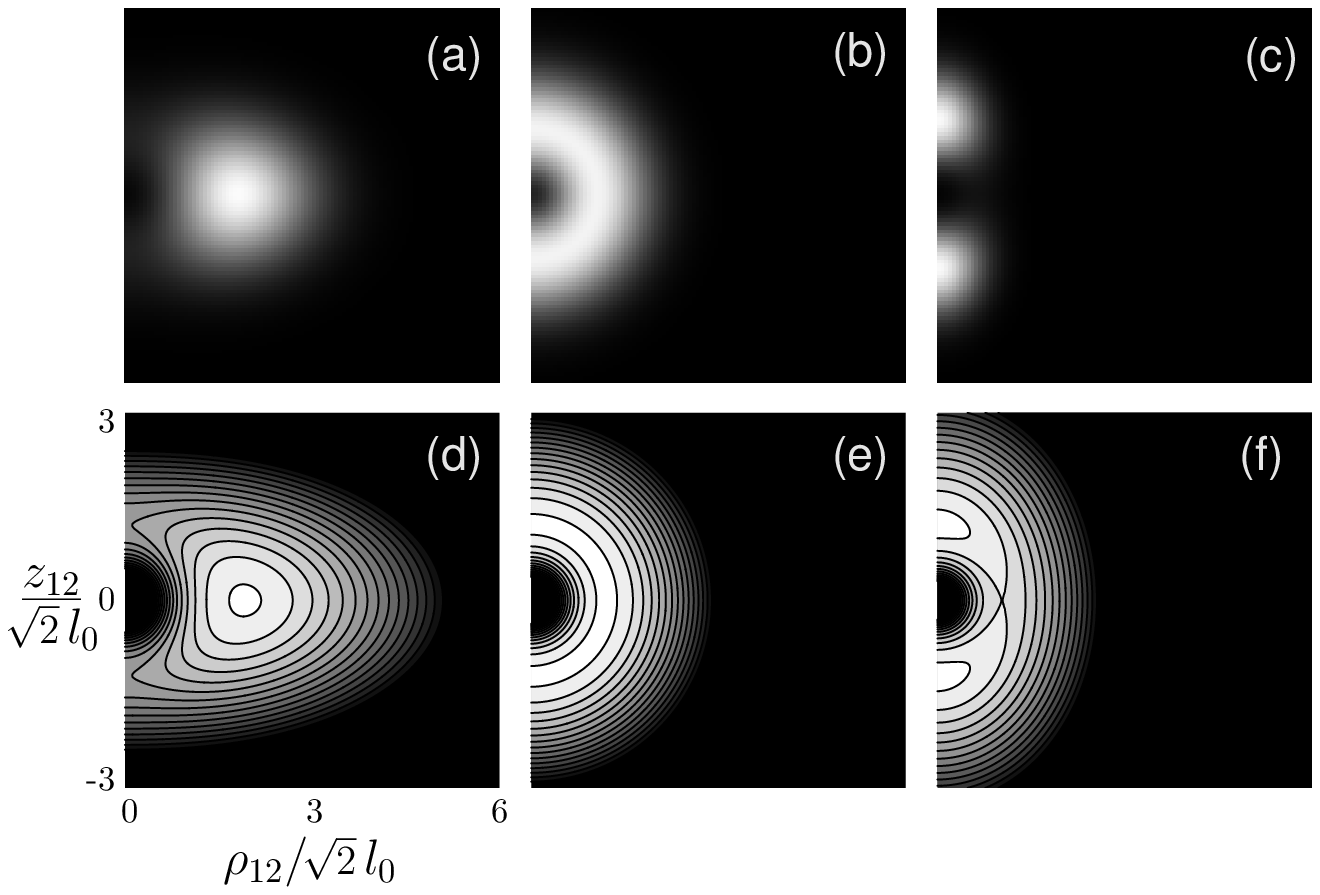}
\end{center}
\caption{\label{fig2}
The probability density $\vert\psi({\bf r}_{12})\vert^2$
of the lowest $m = 0$ state (top) and the contour plots of the
potential surface (bottom) for the QD with $\omega_z/\omega_0 = 2$
and $R_W = 10$ shown in ($\rho_{12},z_{12}$)-plane for the cases:
(a,d) $\omega_L/\omega_0 = 0$ ($\omega_z/\Omega = 2$), (b,e)
$\omega_L/\omega_0 = \omega_L^{\rm sph}/\omega_0 = 1.73205$
($\omega_z/\Omega = 1$) and (c,f) $\omega_L/\omega_0 = 2.29129$
($\omega_z/\Omega = 0.8$).} 
\end{figure}

To get deep insight into this transition we calculate the
probability density $\vert\psi({\bf r}_{12})\vert^2$ and potential
surfaces for various values of the magnetic field (see
Fig.~\ref{fig2}). Since the symmetry is exact for any strength of
the electron-electron interaction at the transition point, in
order to illuminate the effect, we use $R_W = 10$. For the
magnetic field $\omega_L < \omega_L^{\rm sph}$ the density maximum 
is located in the $(x_{12},y_{12})$-plane ($z_{12} = 0$,
see Fig.~\ref{fig2}(a)). For $\omega_L > \omega_L^{\rm sph}$,
however, there are two maxima located symmetrically along the
$z_{12}$-axis ($\rho_{12} = 0$, see Fig.~\ref{fig2}(c)).
The analysis of the behaviour of the stationary point of the
potential $V = \frac{1}{2}\mu(\Omega^2 \rho_{12}^2 + \omega_z^2
z_{12}^2) + k/r_{12}$ as a function of the magnetic field provides
the explanation. For $\omega_L < \omega_L^{\rm sph}$ ($\Omega <
\omega_z$) the stationary point $\rho_{12} = \rho_0$, $z_{12} = 0$
is the minimum of the potential surface (see Fig.~\ref{fig2}(d)). Here
$\rho_0 = (k/\mu\Omega^2)^{1/3}$ \cite{PSN}. By increasing the
magnetic field over the value $B_{\rm sph}$ ($\Omega > \omega_z$)
the stationary point transforms to the saddle point and two new
minima appear, divided by a potential barrier (see
Fig.~\ref{fig2}(f)). In other words, for $m = 0$ a bifurcation of
the stationary point located at $(\rho_0,0)$ occurs at the value
of magnetic field when $\omega_L = \omega_L^{\rm sph}$ (see
Fig.~\ref{fig2}(b,e)). In the domain $\Omega > \omega_z$, for $m =
0$, the minima are located at $z_{12} = \pm z_0$ in the
$z_{12}$-axis ($\rho_{12} = 0$), where $z_0 =
(k/\mu\omega_z^2)^{1/3}$. Similar behavior is observed for the states with
$m\neq 0$. However, the effect is less prominent due to weakening of correlations
by the orbital electron motion.

\section{Conclusions}

Summarizing, we have shown that the 3D approach provides a
consistent description  of the shape transition in the excited states of 
two-electron QDs under the magnetic field. The entanglement of the lowest 
state with $m=0$, being first a decreasing function of the magnetic field, starts to
increase after the transition point with the increase of the
magnetic field. 
This behaviour is
understood as the transition from the lateral to the vertical
localization of the two-electron probability density for this
state, which becomes  prominent at strong Coulomb interaction
($R_W \gg 2$).

\ack
This work is partly supported by RFBR Grant No.11-02-00086
(Russia), Project 171020 of Ministry of Education and Science of Serbia,
Spanish MICINN Grant No. FIS2008-00781, and Project FQM-2445 of the Junta
de Andalucia (Spain).

\end{document}